\documentclass{aastex}
\usepackage{emulateapj5}

\newcommand{\cena}{NGC~5128}
\newcommand{\etal}{et al.\ }
\newcommand{\vi}{$V$--$I$}

\shorttitle{Young Blue Tidal Stream in NGC 5128}
\shortauthors{Peng, Ford, Freeman, \& White}

\begin{document}


\submitted{To appear in the December 2002 issue of The Astronomical Journal}
\title{A Young Blue Tidal Stream in NGC 5128}


\author{Eric W. Peng\altaffilmark{1} and Holland C. Ford\altaffilmark{1,2}}
\affil{Department of Physics and Astronomy, Johns Hopkins
        University, Baltimore, MD, 21218, USA}
\email{ericpeng@pha.jhu.edu, ford@pha.jhu.edu}

\author{Kenneth C. Freeman}
\affil{RSAA, Australian National University, Canberra, ACT, Australia}
\email{kcf@mso.anu.edu.au}

\and

\author{Richard L. White}
\affil{Space Telescope Science Institute, 3700 San Martin Drive, 
	Baltimore, MD 21218, USA}
\email{rlw@stsci.edu}

\altaffiltext{1}{Visiting Astronomer, Cerro Tololo Inter-American Observatory,
which is operated by the Association of
Universities for Research in Astronomy, Inc.  (AURA) under cooperative
agreement with the National Science Foundation.}
\altaffiltext{2}{Space Telescope Science Institute, 3700 San Martin Drive,
	Baltimore, MD 21218, USA}


\begin{abstract}
Stellar streams in galaxy halos are the natural consequence of a
history of merging and accretion.  
We present evidence for a blue tidal stream of {\it young} stars in the
nearest giant elliptical galaxy, \cena\ (Centaurus~A).  Using optical
{\it UBVR} color maps, unsharp masking, and adaptive histogram
equalization, we detect a blue arc
in the northwest portion of the galaxy that traces a partial ellipse with an
apocenter of 8~kpc.  We also report the discovery of numerous young star
clusters that are associated with the arc.  The brightest of these
clusters is spectroscopically confirmed, has an age of $\sim350$~Myr,
and may be a proto-globular cluster.
It is likely that this arc, which is distinct from the surrounding shell
system and the young jet-related stars in the northeast, is a
tidally disrupted stellar stream orbiting the galaxy.  Both the age
derived from the integrated optical colors of the stream and its
dynamical disruption timescale have values of 200--400~Myr.  
We propose that this stream of young stars was 
formed when a dwarf irregular galaxy, or similar sized gas fragment,
underwent a tidally triggered burst of star formation as it fell into
\cena\ and was disrupted $\sim300$~Myr ago.  The stars and star clusters
in this stream will eventually disperse and become part of the main body
of \cena, suggesting that the infall of gas-rich dwarfs plays a role in the
building of stellar halos and globular cluster systems.
\end{abstract}




\section{Introduction}

The accretion and subsequent tidal disruption of low mass galaxies
is suspected to be an important agent in the evolution of galaxy halos.
Both the outer halo of the Milky Way (Searle \& Zinn 1978) and the
globular cluster systems of nearby ellipticals (C\^{o}t\'e, Marzke, \&
West 1998) may have been formed
by the late infall of low mass fragments.
Hierarchical models of galaxy formation suggest that structure first
forms on small scales, and later combines to form larger galaxies
(e.g. Klypin et al.\ 1999).
In our own Galaxy, the discovery of tidal streams associated with the 
Sagittarius dwarf galaxy (Ibata et al.\ 2001a) as well as tidal tails
around Galactic globular clusters (Odenkirchen et al.\ 2000) 
give credence to the idea of
``spaghetti halos''---that galaxy halos may be comprised of the many
remnants of tidally disrupted dwarf galaxies (Morrison et al.\ 2000).

The study of streams in galaxy halos is valuable, both for
understanding the stellar mass assembly of galaxies, and as probes of
the galaxy potential (Johnston, Sackett, \& Bullock 2001, hereafter JSB01).  
However, the identification of accretion
events that still maintain their spatial coherence (ages of less than a
few Gyr) is challenging.  In our own Galaxy, debris streams have been
identified using star counts and kinematics (Majewski et al.\ 1999; Helmi
et al.\ 1999; Newberg \etal 2002).  
In external galaxies, these streams can only be detected
with deep imaging at surface brightness levels fainter than
27~mag/arcsec$^2$.  Low surface brightness features in many galaxies,
which may or may not be debris trails, were detected by Malin \&
Hadley (1997) using photographic amplification techniques. 
Until now, perhaps the best example of an observed tidal streamer is 
the one in NGC~5907 (Shang et al.\ 1998).  There is also strong evidence
that tidal streams exist in the halo of M31 (Ibata \etal 2001b; Choi,
Guhathakurta \& Johnston 2002).  However, in all cases outside the Local Group,
distance and low surface brightness preclude any direct study of the
stellar populations and kinematics of the accreted stars.

\begin{figure*}[t]
\epsscale{1.9}\plotone{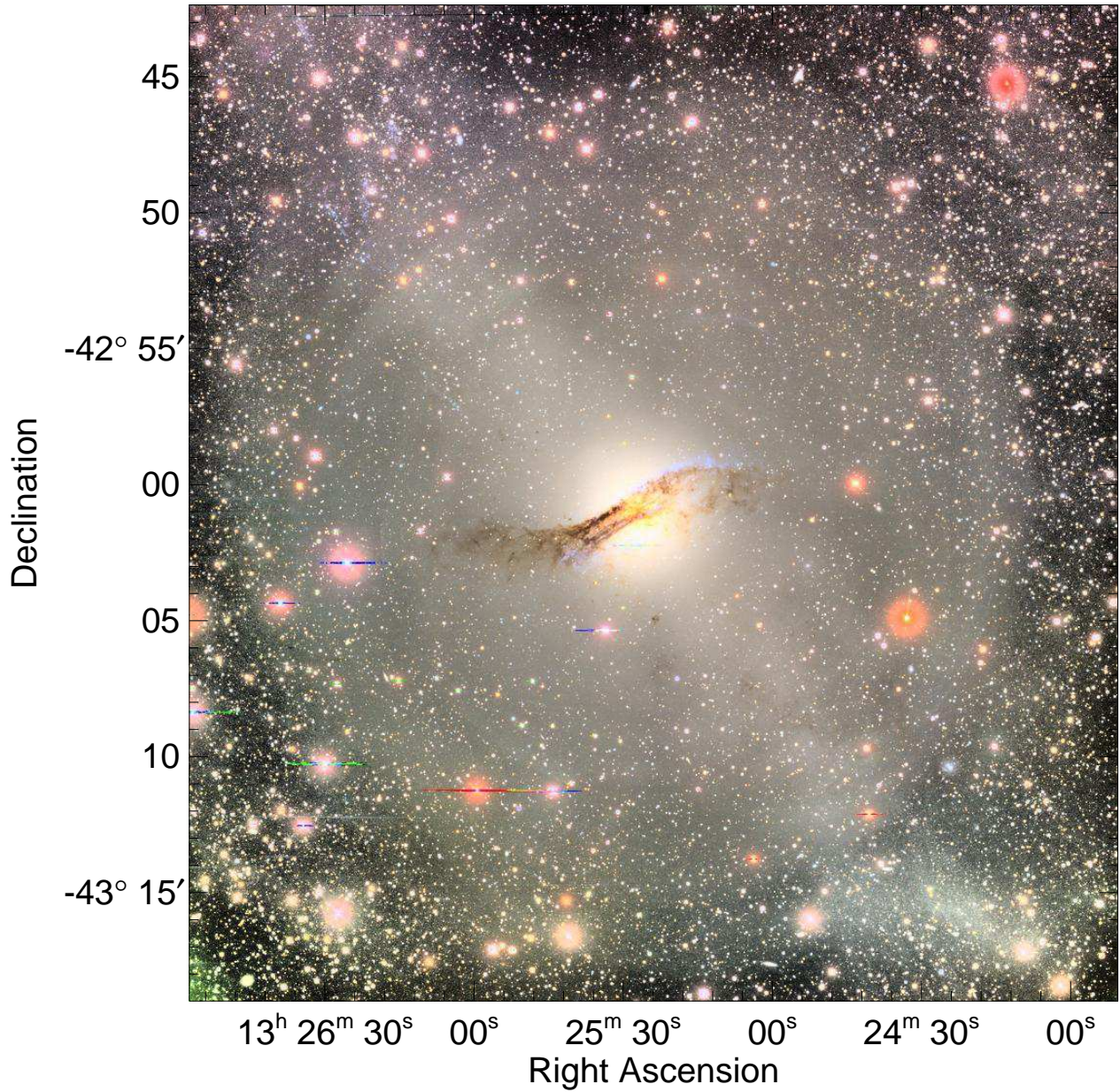}
\caption{A {\it BVR} color image of \cena\ created by applying our AHE
algorithm to our Mosaic observations.  
This image was processed using the modified
adaptive histogram equalization (AHE) method described in
section~\ref{sec:ahe}.  Colors assigned to pixels were
determined from flux ratios in the original images.  
The AHE processing allows one to see that the
halo of the galaxy fills most of the frame, extending well beyond the
familiar dust lane region along the photometric minor axis.  Small
gradients and patchiness in color are due to sky variations and
systematic flat-fielding errors at the $\sim1$--2\% level.  In all
figures, north is up and east is to the left.
\label{bvrimage}}
\end{figure*}

\section{NGC~5128 (Centaurus~A)}

NGC~5128 (also the radio source Centaurus A) is the nearest easily
observable giant elliptical galaxy, and is the prototype for a
post-merger elliptical.  A {\it BVR} color image of \cena\ is shown 
in Figure~\ref{bvrimage}.  The prominent central rotating disk of gas 
and dust (Graham 1979), optical shells
(Malin, Quinn, \& Graham 1983, hereafter MQG83), and \ion{H}{1} shells 
(Schiminovich et al.\ 1994) all point to a merger event within the last
Gyr.  It is also the prototype Fanaroff-Riley Class I radio galaxy, with
an AGN-driven radio jet believed to be inducing star
formation in the northeast halo regions (Fassett \& Graham 2000).
While \cena\ appears to be a uniquely complex system, it is only because
of its proximity that we are able to discern and study features that are
likely to be common in more distant galaxies.  We adopt a
distance of 3.5~Mpc ($m-M=27.72$) to NGC~5128 given by the planetary nebula
luminosity function (Hui et al.\ 1993).  This distance is consistent
with measurements from the globular cluster luminosity function
(G.Harris \etal 1984) and the tip of the
red giant branch (Soria \etal 1996; Harris \& Harris 2000). 
At this distance, 1~kpc~$\sim 1\arcmin$, and the full extent of
the galaxy's halo covers over two degrees of sky.
NGC~5128's effective radius is
given by Dufour et al.\ (1979) to be $305\arcsec$, or 5.2 kpc.
Throughout this paper, we adopt 
($13^{\rm h}25^{\rm m}27\fs6$, $-43\degr01\arcmin09\arcsec$, J2000) 
as the position of the center of NGC~5128.

Studies of the NGC~5128 halo regions are difficult because of its large
extent on the sky and its relatively low Galactic latitude
($+19\degr$).  Nevertheless, it is an extremely valuable case study
because its proximity facilitates resolved stellar population studies,
and its post-merger state makes it a likely
candidate for harboring recent accretion and disruption events.
We are conducting an optical study of the halo of NGC~5128 that
encompasses the planetary nebulae, globular clusters, and unresolved
stellar light.  In this paper, we discuss our search for tidal streams
in NGC~5128's halo. 

\section{Observations and Data Reduction}

We obtained our observations during 2000 June 1--4 at the 4-meter
Blanco telescope of the Cerro Tololo Inter-American
Observatory (CTIO).  We observed three fields in NGC~5128 with the Mosaic II
optical CCD camera, which has $0\farcs26$ pixels and 
a 0.5-degree  field of view.
In this paper, we discuss our central pointing which includes the
inner $\sim4 r_e$ of the galaxy.  

We imaged the galaxy through the Johnson-Cousins {\it UBVRI} filters.  Total
effective exposure times were 3600s, 1500s, 1800s, 1000s, and 1000s,
respectively.  Conditions were moonless and photometric for
the entirety of the observing run, with a typical seeing of 1\arcsec.  
We split our observations into a series of
five dithered exposures in order to reject cosmic rays and fill in
the gaps between the CCDs.  
We observed standard stars from Landolt (1992) and obtained photometric
solutions in {\it UBVRI} with rms errors of 0.07, 0.02, 0.02,
0.02, and 0.03 magnitudes, respectively.

We reduced the data using MSCRED, a software package for IRAF that is
specifically designed for the reduction of data from NOAO Mosaic CCD
cameras (Valdes 1998).  Images were bias subtracted, flat-fielded using 
both dome and night sky flats, astrometrically regridded, and combined
with a 3-sigma upper threshold rejection of cosmic rays.  
Because the galaxy filled most of the field of view, the individual CCDs
were not background subtracted before combining.

\section{Image Processing}
Some of the best work to date on faint halo features in nearby galaxies 
was done by Malin (1978) with photographic plates of \cena.
It is only with the relatively recent development of wide-field mosaic
CCD cameras that we can apply modern digital processing techniques to 
a comparable field of view.  Below, we describe the techniques that we
used to bring out faint or low-contrast features.

\subsection{Adaptive Histogram Equalization}
\label{sec:ahe}

Astronomical images often contain vastly more information than can be
displayed in a conventional grayscale or color representation of the
image.  Observers will be familiar with the process of fiddling with
the contrast and brightness of an image in order to look for features
buried in a strongly varying background.  This problem is particularly
difficult for our image of \cena\ because the galaxy extends across
the entire field of view and varies in brightness by more than a factor
of 100 from the center to the edge.

Adaptive histogram equalization (AHE; Pizer et al.\ 1987) is an image
processing algorithm that is designed to solve this very problem.
Although AHE has been relatively commonly used for medical images, it
appears to have been used at most rarely in astronomy.  We here
describe the algorithm briefly and discuss some improvements we have
made.

The more familiar (non-adaptive) histogram equalization algorithm for a
grayscale image maps the input image pixel values monotonically to the
output such that the output image has an approximately equal number of
pixels at each gray level.  The histogram of pixel values for a typical
astronomical image is strongly peaked around the sky brightness; the
histogram-equalized image has a flat distribution of pixel values and
therefore uses most of the gray levels near the sky.  Such an image is
typically much better for seeing faint structures than is a linear
grayscale mapping.

The basic algorithm for AHE is straightforward: one applies standard
histogram equalization to a {\sl region} around each pixel and chooses
a {\sl local} gray level mapping that is determined by the distribution
of input pixel values in that region.  The contrast is allowed to vary
across the image, with the scale of variation determined by the region
size (which is a parameter of the method.)  Pizer et al.\ (1987) discuss
methods of performing the computation and propose a simple algorithm
where the grayscale is computed for non-overlapping tiles covering the
image and then is linearly interpolated for pixels not in the centers
of tiles.

Both histogram equalization and AHE have the drawback that they can
unacceptably enhance the noise in images. For example, in a typical
astronomical image 99\% of the pixel values fall within 3 sigma of the
sky brightness; if that distribution is made perfectly flat by AHE, the
resulting image looks basically like white noise with a few saturated
regions where objects are found. Pizer et al.\ have addressed this
problem by clipping the intensity histogram before determining the gray
scale to limit the maximum contrast in the image.  The clipped parts of
the histogram are uniformly distributed into the unclipped bins of the
histogram.  This clipped histogram equalization is equally applicable
in the adaptive and non-adaptive cases.  In their approach there is a
single parameter specifying the maximum slope of the function mapping
input to output pixel values.

\begin{figure*}[t]
\epsscale{1.9}\plotone{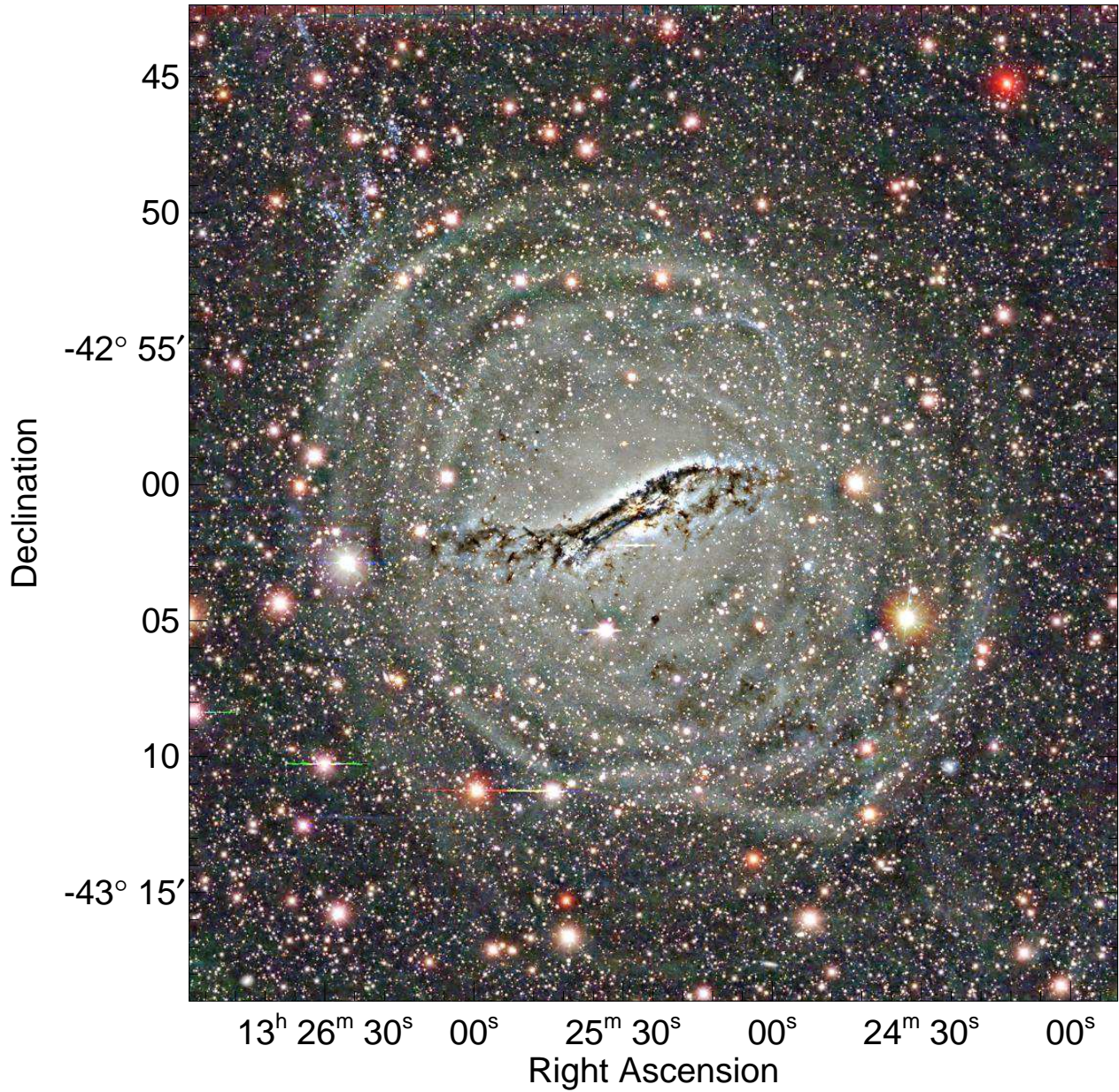}
\caption{A {\it BVR} color image of of \cena, which was 
also processed using the modified
AHE method.  In this case we chose to emphasize structure on smaller
scales.  Colors were assigned based on flux ratios in
the AHE-processed images rather than in
the originals in order to emphasize low-contrast color features.
Faint dust extensions are visible along the major axis, especially to
the southwest, as is extensive halo shell structure.  A blue elliptical
arc is visible at approximately ($13^{\rm h}25^{\rm m}$,
$-42\degr55\arcmin$).
\label{ahe_full}
}
\end{figure*}

We have made a few modifications to the Pizer et al. AHE algorithm to
produce better results for astronomical images:
\begin{enumerate}
\item In the clipping algorithm, we incorporate a CCD noise model
    (readout noise plus Poisson counting noise); the clipping parameter
    is then specified as the maximum fraction of output gray levels
    that can be assigned within any 1-sigma intensity range.  This
    gives good results and also determines the natural bin size for
    floating point images.  Our algorithm thus works on either integer
    or floating point images.

\item In computing the lookup tables for the adaptive regions, we find it
    necessary to use overlapping tiles (shifting by less than the tile
    size) in order to eliminate visual artifacts when the tiles are
    small.  Typically we sample the histogram at an interval of half
    the region size, although for smaller regions the sampling interval
    may need to be one third.  Pizer et al. discuss this approach but
    conclude it is not needed for their images.

\item We do not necessarily fix the maximum and minimum values for the
    histogram across the entire image, instead allowing the values to
    vary depending on the range of pixel values in that region.  This is
    needed for high dynamic range images.
\end{enumerate}


\begin{figure*}[t]
\epsscale{1.7}\plotone{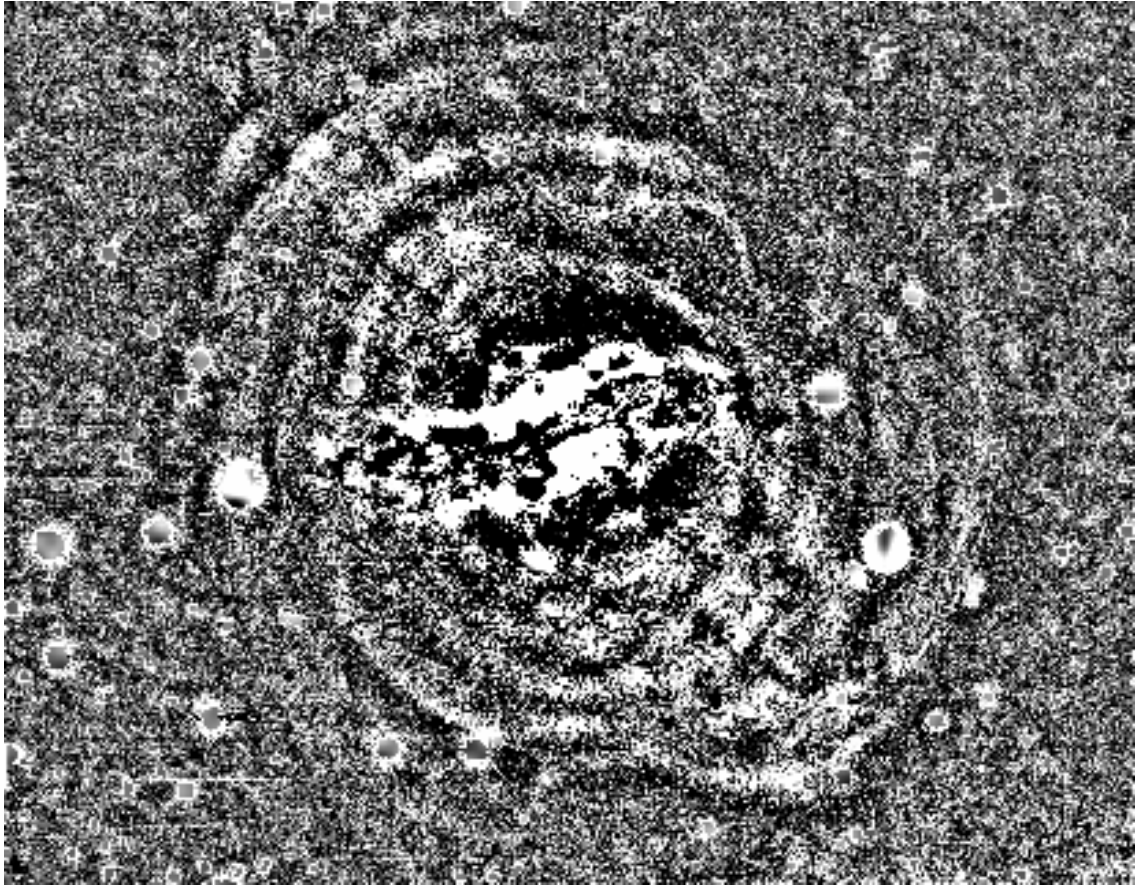}
\caption{Shell system of \cena.  This is our {\it V}-band image
after being processed using the technique of unsharp masking.  The field
of view is $32\arcmin\times25\arcmin$, centered on \cena.  Many
faint shells are visible, even in the inner regions.  
Circular artifacts are due to the masking of bright stars.
\label{shells}
}
\end{figure*}


To apply AHE color (multi-band) images, we have used two approaches.
The first is to perform AHE on a single combined image from the three bands,
and then to create a color output image using the AHE image for the
intensity and the ratios of the input image pixel value to determine the RGB
values of the output.  This has the advantage that it preserves the
colors of the input pixels while enhancing the contrast.  A color
AHE version of our \cena\ image created using this technique
is shown in Figure~\ref{bvrimage}.  Note that it shows detail across the whole
range of the image, from the faint outer reaches of the galaxy (where
the shells and jet-induced star formation regions are visible) to
the center of the galaxy (where details of the dust lane can be
seen.)  The colors here are consistent and comparable across the image;
gradients in the apparent colors of stars across the field are
the result of slight flat-fielding inconsistencies.

The second approach for applying AHE to color images is to
independently equalize the three bands and then to combine them.  This does
not generate output colors that map directly to the input colors.
However, if one is interested in the presence of a low-contrast blue
feature on a bright red background, this second approach will do a
better job of bringing out the contrasting color of the weak features.
Figure~\ref{ahe_full} shows a color AHE version of our image using this second
method.  (For this image, we have also adaptively filtered each
filter's data using a wavelet shrinkage method as suggested by
Donoho (1992)). The difference from Figure~\ref{bvrimage} is striking.  
In this image the complex system of shells is clearly visible, as is a
blue elliptical arc in the northwest quadrant of the galaxy.  

\subsection{Object Detection and Masking}

Most of the discrete sources visible in the field of \cena\ are
foreground stars in the Milky Way.  Because we are only interested in
the unresolved stellar light of \cena, another way to better reveal the
structure of the galaxy is to identify and interpolate across
foreground and background objects.

We used the object detection package SExtractor (Bertin \& Arnouts 1996)
to create a catalog with positions and magnitudes for all objects.
Detection and 
aperture definition were done on the {\it V}-band image, which went as
deep or deeper than images taken through the other filters.  We used SExtractor
because it does a good job of fitting the
variable background introduced by the underlying galaxy light, and also
because we wished to measure magnitudes for globular clusters, which are
slightly resolved.  In order
to create an image free of discrete sources such as foreground stars,
globular clusters, and background galaxies, we used the pixel mask
created by SExtractor to identify the location of ``object'' pixels.
We replaced these pixels with the estimated background values as
determined by the package's mode and median algorithm.  This is
a valid approximation where the galaxy's light varies slowly and on
large spatial scales, and allows us to create a ``source-free'' image.  

\begin{figure*}[t]
\plotone{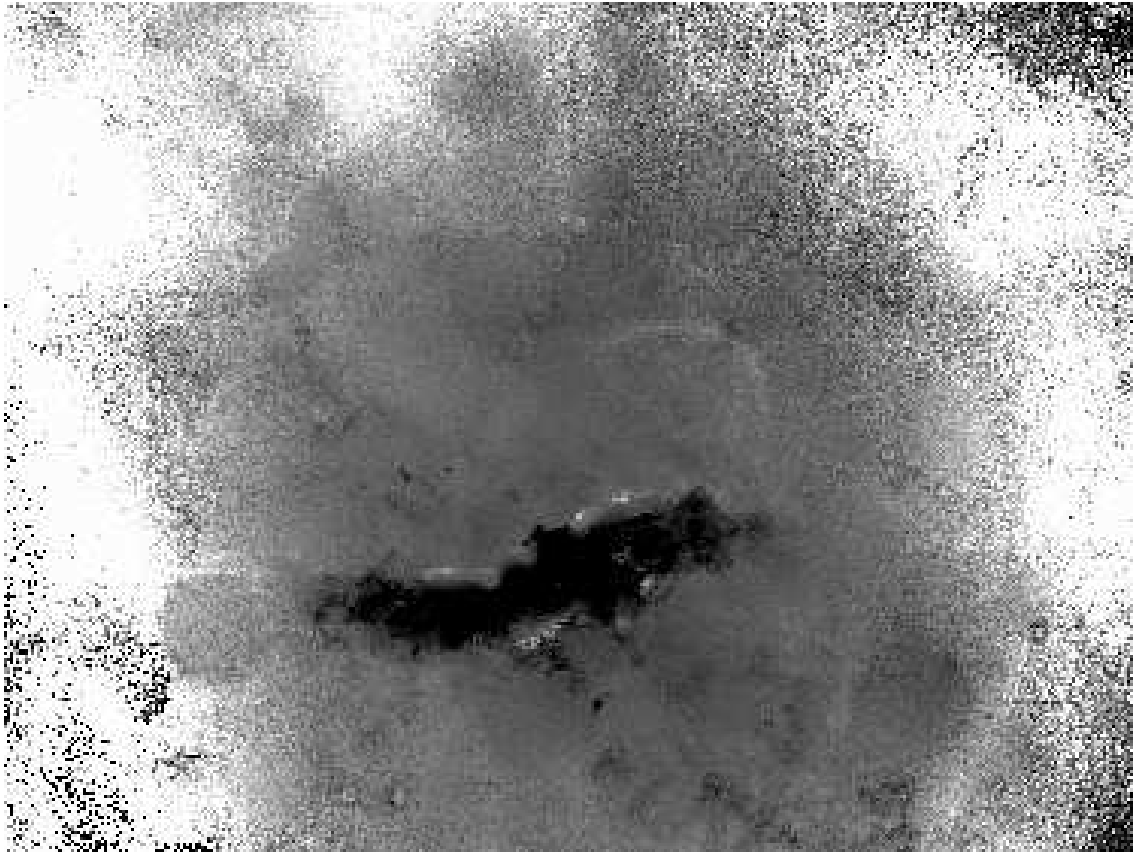}
\caption{($B$--$R$) color map of a $32\arcmin\times24\arcmin$ region of
\cena.  The field of view is similar to Figure~\ref{shells}, but
translated north by $4\arcmin$.  This map was produced by
dividing the $B$ image by the $R$ 
image after object masking and sky subraction.   The map was smoothed by
a $100\times100$~pixel median filter.  Circular artifacts are
remnants of masking of bright stars.  Darker shades are redder.  As
expected, the main body 
of the galaxy and the central dust disk is red.  However, there exists a
distinctly blue elliptical arc whose brightest portion is $8\arcmin$
northwest of the galaxy center.  Of the many shells that are visible in
the unsharp masked image (Figure~\ref{shells}), 
only one in the southwest corner shows any
contrast in this color map.
\label{colormap}}
\end{figure*}

\subsection{Unsharp Masking}

Unsharp masking is an image processing technique that emphasizes sharp
features on a smooth background.  By subtracting or dividing out a
smoothed version of the image from the original, one can increase the
contrast of features that vary on scales smaller than the smoothing box.
Malin and Carter (1983) successfully applied unsharp masking to
photographic plates to reveal faint shells and filaments in galaxies.  

NGC~5128 is a known shell galaxy, having had some of its shells
photographically cataloged in pioneering work by MQG83. We unsharp
masked our CCD images to look for more fine structure in \cena.  
We created a blurred image of the galaxy by applying a $300\times300$
pixel median filter to the source-free image.  We subsequently
subtracted the blurred image from the source-free image.
The final processed {\it V}-band image is shown in Figure~\ref{shells}. 
With modern CCD data, we have the dynamic range and areal coverage 
to capture the richness of the \cena\ shell system from its bright central
regions to its faint halo.  Figure~\ref{shells} shows that 
\cena\ possesses many shells, some of which have not been previously
cataloged.  Also visible along the major axis in the NE are two faint radial
extensions that may be reflected light from the embedded AGN.
We defer discussion of these and other features in our 
CCD fields to a later paper.

\subsection{Color Maps}

Faint stellar features that are significantly younger or
more metal-poor than the bulk of the galaxy will be bluer than their
surrounding light, and as a result will stand out in a color map.
Because some objects that may have recently merged with a giant
elliptical---like spirals and dwarf galaxies---are generally bluer than
ellipticals, searching for anomalously blue features can be an
efficient way to map recent mergers.  

We created color maps of NGC~5128 using our {\it UBVR} images.  The
varying sky in the {\it I}-band image made it unsuitable for faint
surface photometry.  We first sky subtracted each of the source-free
images, a task that is non-trivial
because the galaxy itself fills most of the CCD frame.  When planning
our observing, we examined a deep photographic image by Malin to be
sure to include in each field a region that is primarily
``sky''---i.e.\ a region
that appears to be free of obvious galaxy light or Galactic cirrus down
to the surface brightness limit of Malin's image 
(29 B mag/arcsec$^2$, Malin 1978).
This darkest portion of our image was used as the reference sky level,
which was then subtracted off for each image.  Because there are so few
true sky pixels in the image, we made the necessary assumption that
the sky level is constant across the image.  After sky subtraction,
we divided one image into another to create color maps.  

\begin{figure*}[t]
\epsscale{1.9}\plotone{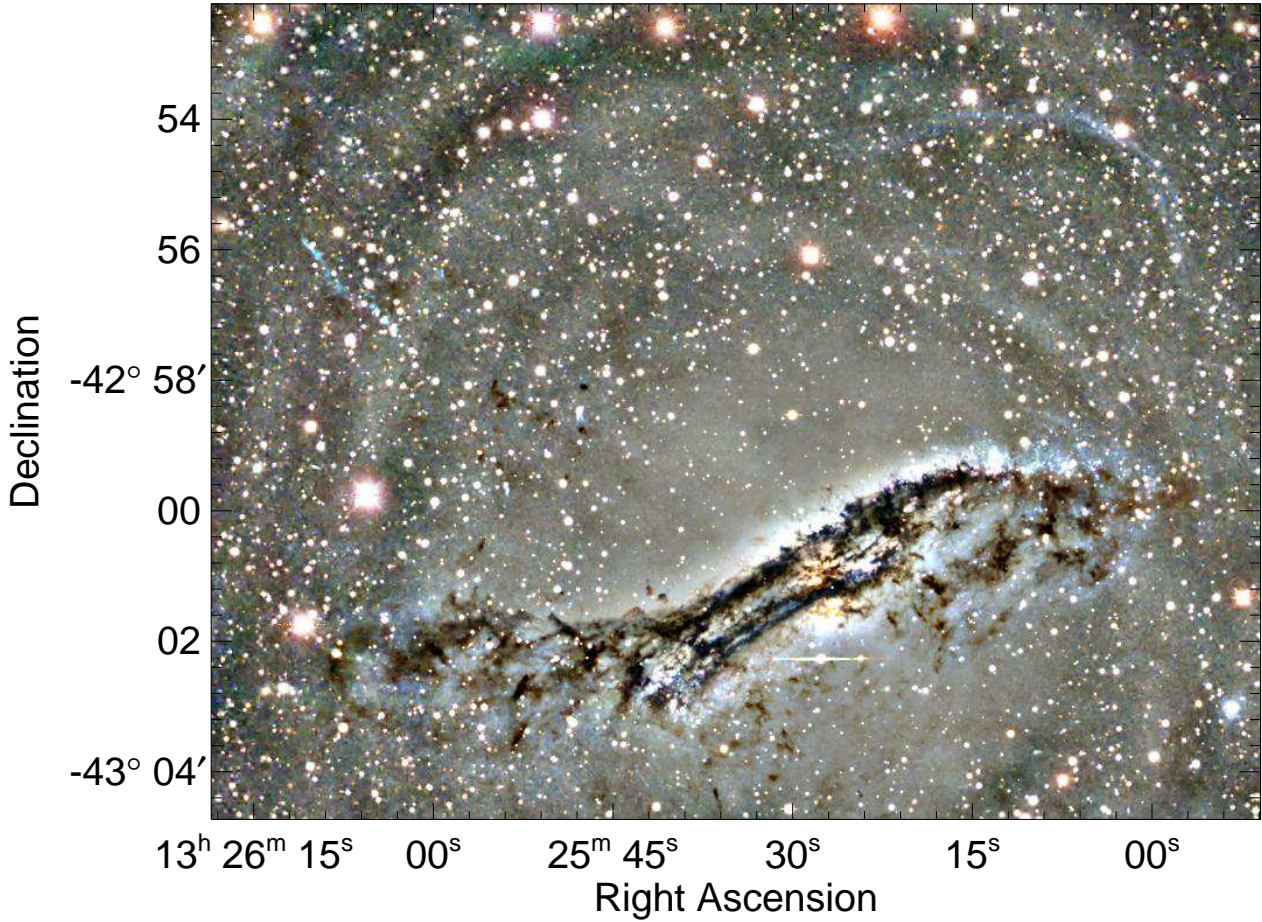}
\caption{A subsection of Figure~\ref{ahe_full} that shows the blue
arc and dust lane region.  The blue arc is in the upper right quadrant
of the image and appears to have a braided appearance.
\label{ahe_sub}}
\end{figure*}

A smoothed color map with the best
combination of contrast and signal-to-noise, ($B$--$R$), is shown in
Figure~\ref{colormap}.  In this image, light shades represent bluer values,
and dark shades are redder.  The prominent central dust lane is visible
as a dark band, and other patches of dust are also identifiable
along the major axis of the galaxy.  The main body of the galaxy is red,
as one would expect for a typical elliptical.  
An exception to this is the arc-like feature in the
NW portion of the inner halo.  This feature is much bluer than the
surrounding stars, and appears to trace out a partial ellipse around the
galaxy center.  

This arc is also evident in the upper right quadrant of the color image in
Figure~\ref{ahe_sub}, a closeup taken from the
AHE-processed Figure~\ref{ahe_full}.  We note that in this color image,
the arc has a braided appearance.  Furthermore, by applying AHE to a
($B$--$R$) color map, we can enhance features that are both sharp and
blue.  An example of this is shown in Figure~\ref{ahe_cmap}, where we
see that the arc extends southeast of the brightest portion, across the major
axis of the galaxy, still northeast of the galaxy center.

In the following sections, we discuss the properties,
stellar content, and possible origin of this blue arc.
 
\section{The Blue Arc}

\subsection{Geometry}

\begin{figure*}[tp]
\epsscale{1.42}\plotone{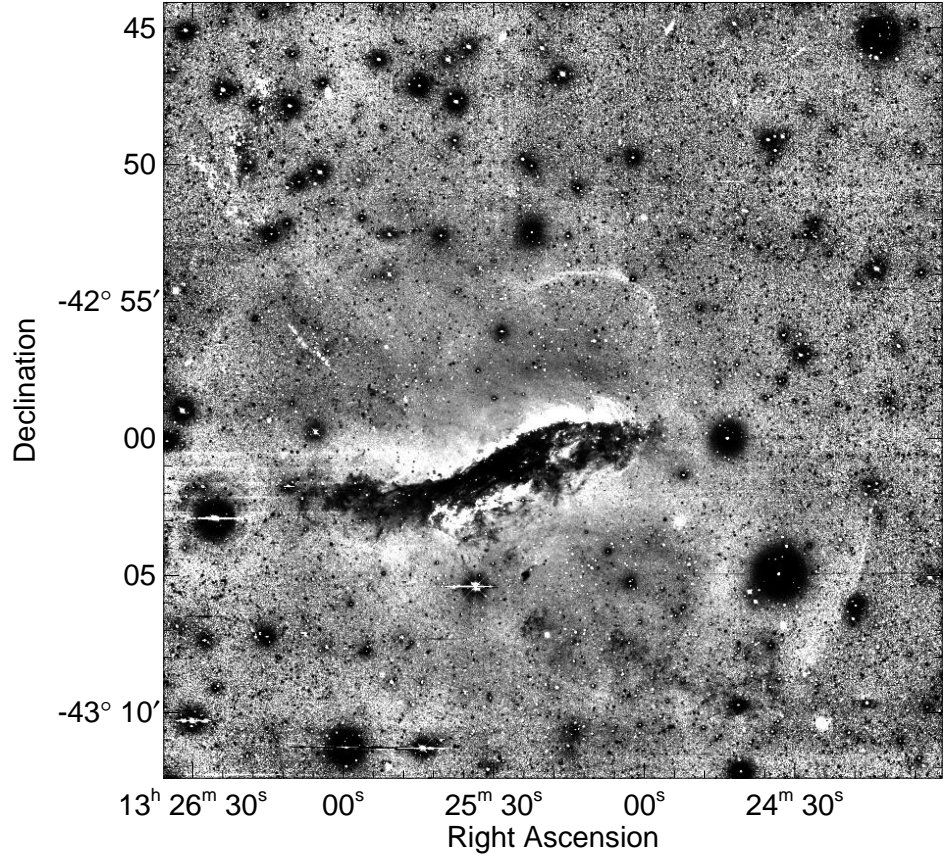}
\caption{A ($B$--$R$) color map processed
with the adaptive histogram equalization method.  This processing was
used to bring out relatively sharp, blue features in the color map.
More evident in this figure is the fragmented and longer extent of the
blue arc across the major axis, as well as a sharp blue shell in the
southwest.  Circular artifacts are due to bright (red) foreground stars.
\label{ahe_cmap}}
\end{figure*}

\begin{figure*}[bp]
\plotone{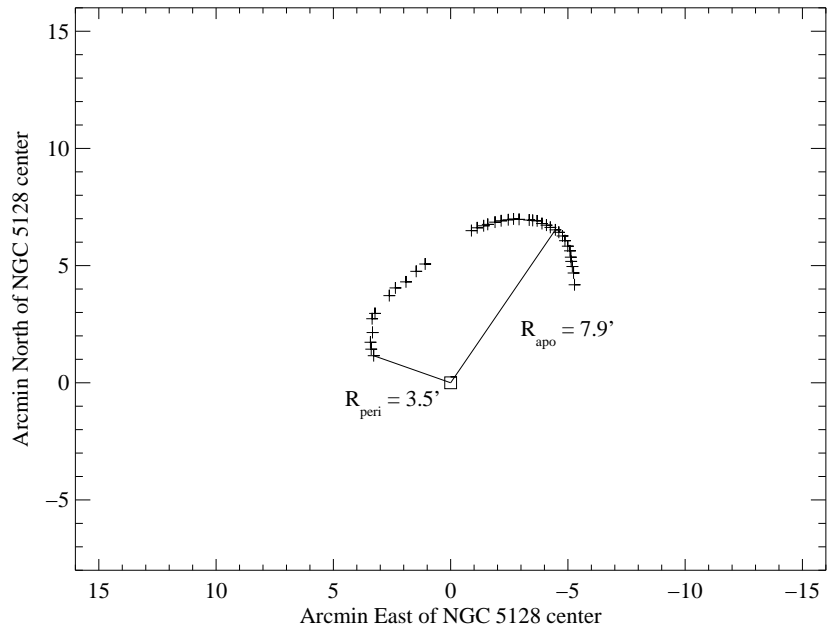}
\caption{Schematic of points along blue arc.  Axes are arc minutes east
and north of the center of \cena, which is labeled by a box at (0,0).
The field of view shown matches that of Figure~\ref{colormap}.  
Crosses label positions of blue light seen in the color map.  
The two lines drawn are the observed apocenter and pericenter for the
tidal stream.
\label{schematic}
}
\end{figure*}

Photometry of faint, low surface brightness features on a variable
background is challenging.  The color maps in Figures~\ref{colormap} and
\ref{ahe_cmap} show
that the arc's bluest portion is in the northwest quadrant of the
galaxy, with less prominent extensions closer in across to the north and
northeast regions.  At these smaller distances from the center, the 
increasing surface brightness of the galaxy and the increased extinction
from dust within \cena\ make the arc more difficult to follow.  We show a
schematic of the traceable regions of the arc in Figure~\ref{schematic}.
The arc's apocenter ($R_{apo}$), the largest projected distance from the galaxy
center, is at 8.1 kpc.  The closest visible region, which we will refer
to as pericenter ($R_{peri}$), is at 3.5 kpc.  The width of the arc at
apocenter is $w = 0.5$ kpc.
Only after determining the arc width did we smooth the color map to
increase the signal-to-noise, using a comparably sized
($100\times100$~pixel) median filter. 

\subsection{Photometry}
\label{photometry}

To measure the light intrinsic to the arc, we first minimized the
varying galaxy background by fitting and then subtracting a model of the
galaxy.  We created this model using the ISOPHOTE package in IRAF by
fitting ellipses to a masked image of the galaxy in each band
(Jedrzejewski~1987).  The dust
lane, blue arc, and all resolved sources were masked and not included in
the fits.  We used the color map to define an arc aperture that
includes the brightest region at apocenter.  We also define two flanking
apertures to measure the background level both inside and outside the arc.  The
background level used and subtracted from the arc aperture was the average
of the values in the two flanking apertures.  The resulting photometry
of the arc is presented in Table~\ref{arcphot}.  All values have been
corrected for Galactic foreground reddening using the extinction maps of
Schlegel, Finkbeiner, \& Davis (1998).  Because the photometry
was measured on the image with all resolved sources already removed, any
light from resolved sources that belong to the arc must also be
included.  We discuss this in the following section.


\begin{figure*}[t]
\epsscale{1.9}\plottwo{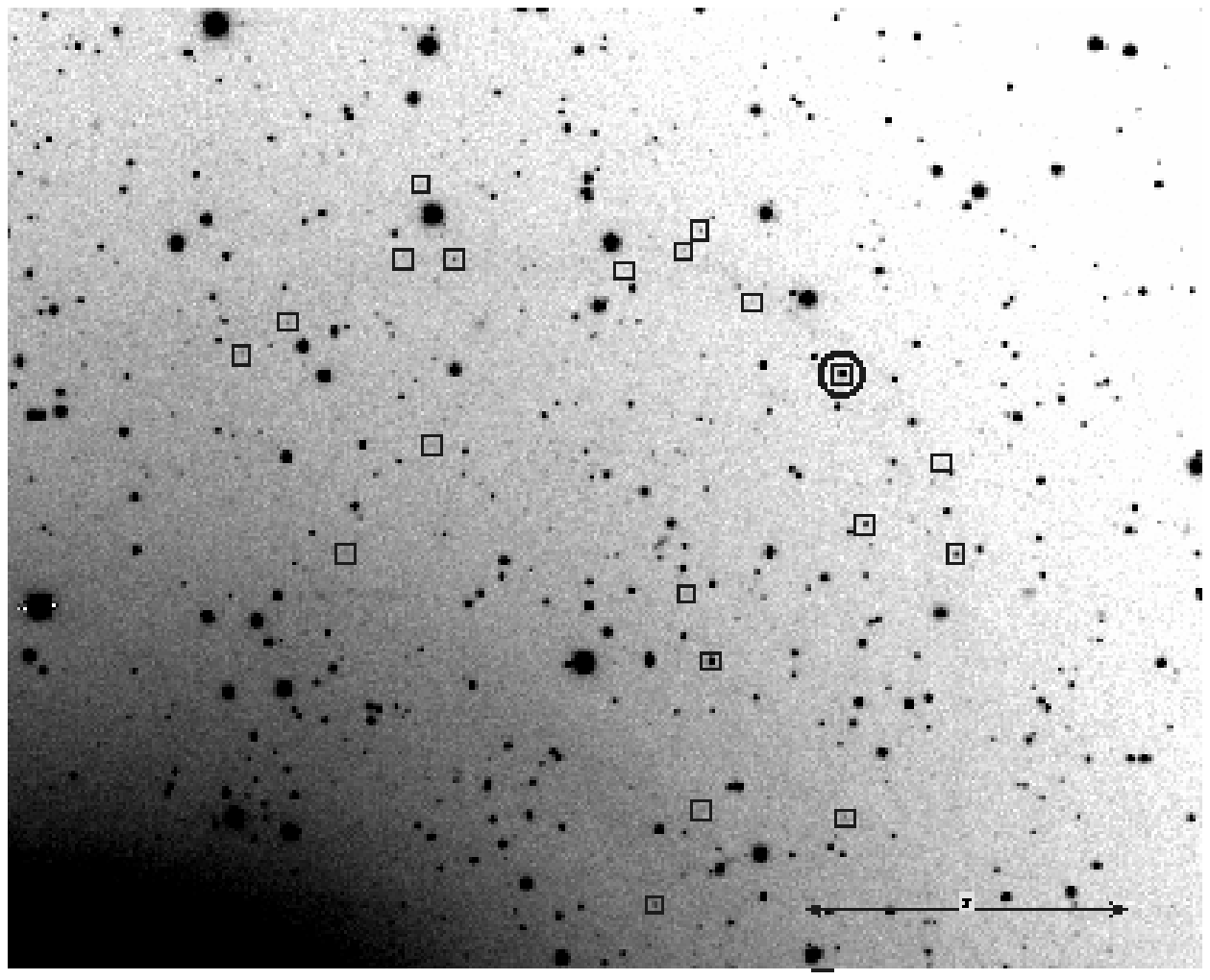}{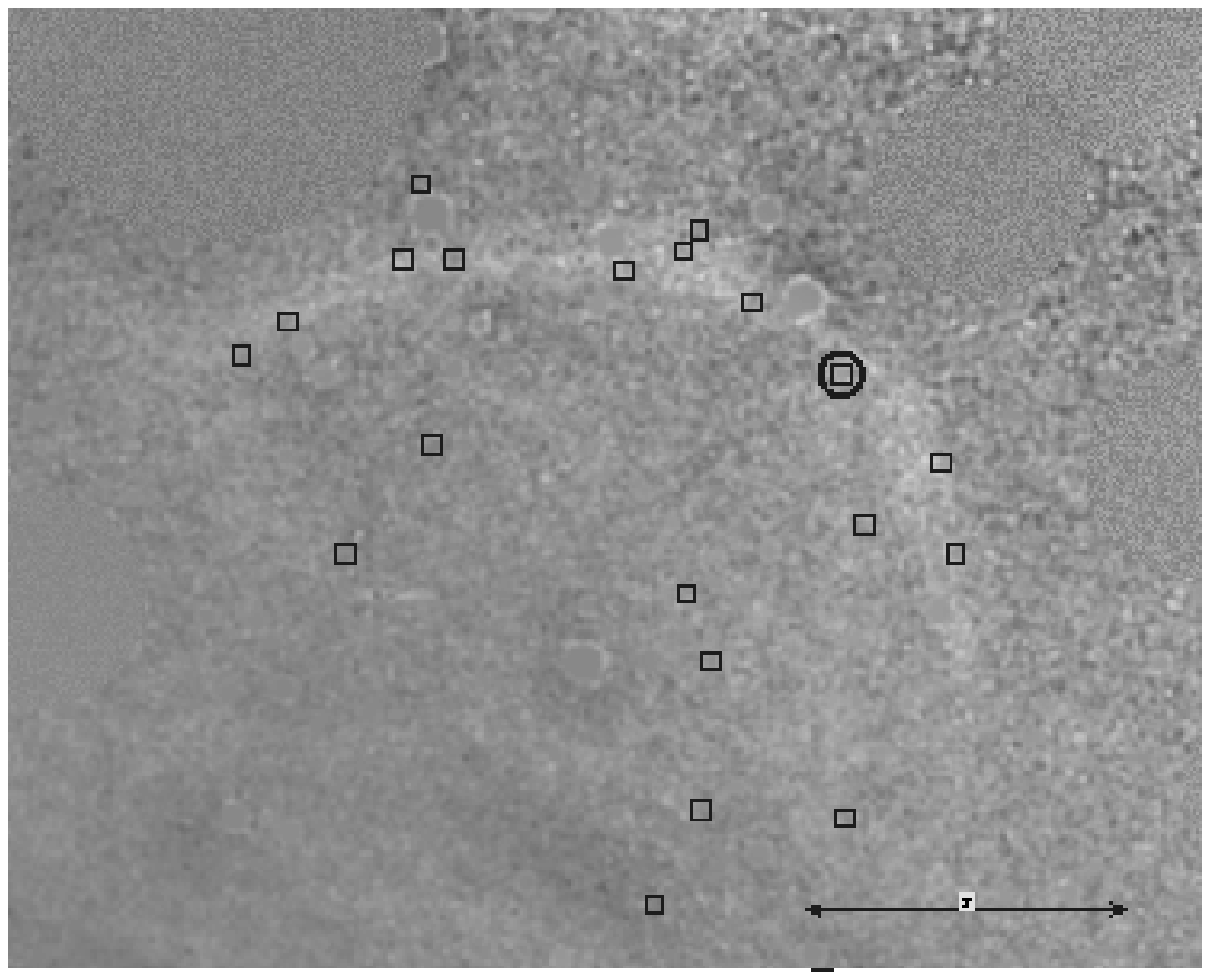}
\caption{Blue Objects and Young Star Cluster along arc in NGC~5128.
These objects were selected as discrete
sources that have apparent magnitudes $18<V_0<23$ and colors bluer than 
(\bv)$ < 0.3$.  To illustrate that many of these lie along the arc,
their locations are plotted as boxes
on both the $B$ image ({\it left}) and the
($B$--$R$) color map ({\it right}).  The spectroscopically confirmed
star cluster is also labeled with a large circle.  The bar in the lower
right corner of each image is $2\arcmin$ long.
\label{clusterpos}}
\end{figure*}

\subsection{Young Star Clusters}

Star clusters are single-age, single-metallicity stellar populations,
and are useful markers of a galaxy's star formation history.  At the
distance of \cena, where $1\arcsec\approx17$~pc, some star clusters are
marginally resolved in ground-based seeing.
In the course of our survey for old globular clusters in the halo of
\cena, we obtained fiber spectroscopy of objects that were slightly
resolved in our imaging.  We acquired these spectra with 
2.5 hr exposures on the CTIO/Hydra spectrograph.  The radial
velocities of these objects allow 
us to determine whether they are genuine clusters or foreground
Galactic disk stars.  

\begin{figure*}[tp]
\epsscale{1.6}\plotone{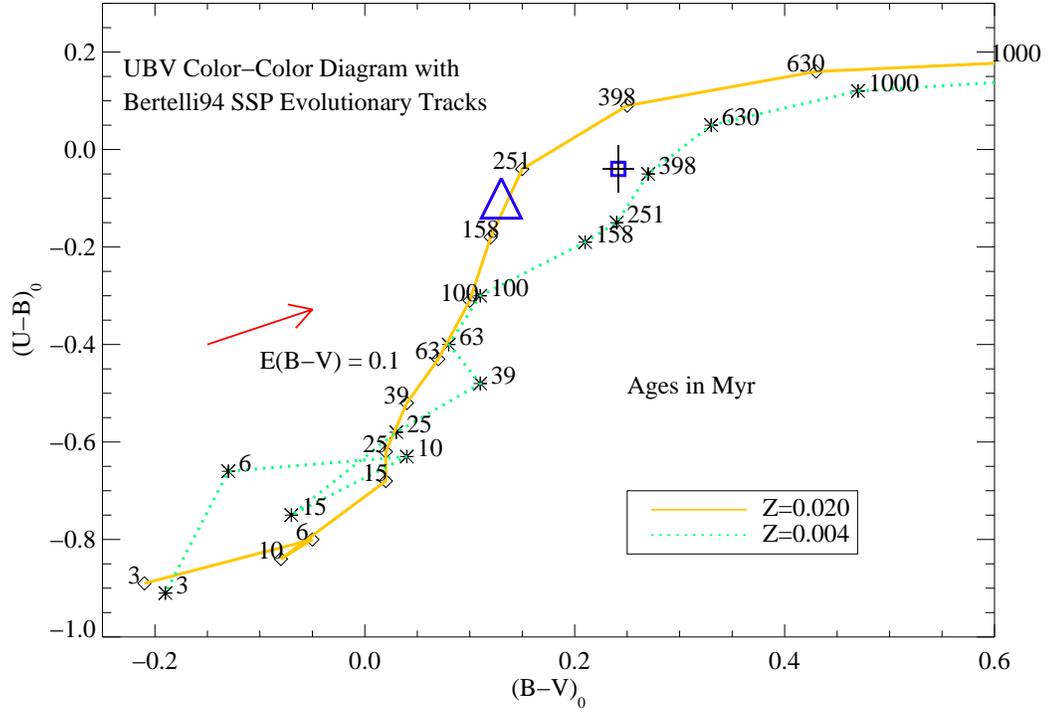}
\caption{{\it UBV} color-color diagram with Bertelli et al.\ (1994)
single stellar population evolutionary tracks.  The solid line and
dotted lines represent metallicities of $Z=0.20$ and $Z=0.04$,
respectively.  The ages of the models as they evolve in color are
labeled in Myr alongside the tracks.  The arrow represents the reddening
vector for $E(B$--$V) = 0.1$~mag.  The square point represents the
colors of the spectroscopically confirmed star cluster, indicating that
it has an age of $\sim350$ Myr.  The triangle represents the integrated
color of the entire arc (from Table~\ref{arcphot}).
\label{bertelli}} 
\end{figure*}

\begin{figure*}[bp]
\epsscale{1.6}\plotone{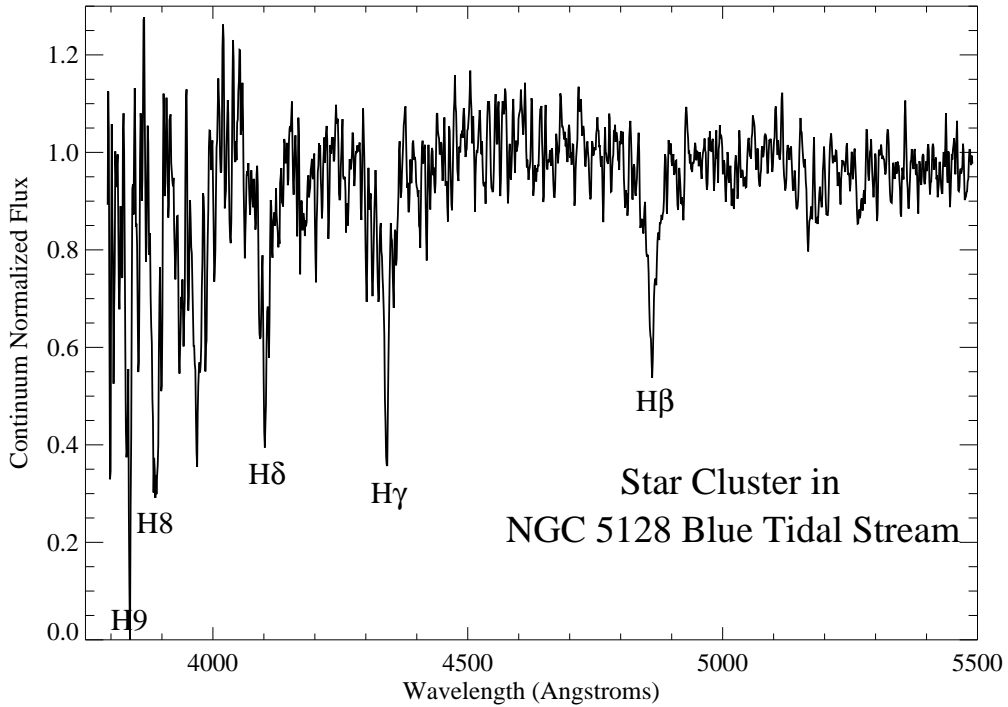}
\caption{Continuum flux normalized spectrum of the young star cluster in the
\cena\ tidal stream.
This cluster was selected for spectroscopy in our globular cluster
program because it was slightly resolved in $1\arcsec$ imaging.  With a
heliocentric velocity of 581~km/s it is a convincing member of the 
\cena\ halo.  The dispersion of the original
spectrum was 1.2\AA\ pixel$^{-1}$ with a resolution of 5\AA, so we
smoothed the spectrum with a 3~pixel boxcar.
The cluster shows strong Balmer absorption, whose equivalent
widths are presented in Table~\ref{balmer}.  
Note that H$\epsilon$ is
typically not measured because it is blended with Ca~H.
\label{clusterfig}}
\end{figure*}

One of these newly confirmed clusters lies directly atop
the blue arc, with its position shown in Figure~\ref{clusterpos}.  
This cluster, which is at ($13^{\rm h}25^{\rm m}01\fs6$, 
$-42\degr54\arcmin40\farcs9$, J2000), 
has a heliocentric velocity of 581~km/s.  This velocity is 
close to the systemic velocity of \cena, $V_{sys} = 541$ km/s (Hui et
al.\ 1995), making it convincingly associated with the galaxy.  
Its {\it UBVRI} broadband colors (Table~\ref{arcphot}) and spectrum show
it to be by far the youngest and bluest of the confirmed clusters in the
\cena\ halo.

Mergers remnants are known to have young or intermediate-age clusters
associated with them (e.g.\ NGC~1316; Goudfrooij \etal 2001).  Because
\cena\ is 
obviously a recent merger, we might expect that some of its star
clusters to be young and associated with the larger merger rather than
directly with the arc.  While it is possible that this blue star cluster
is merely superimposed on the arc, we
believe that this is unlikely.  Our spectroscopic survey included many
objects at similar distances to the galaxy center, including blue
objects that are unresolved, and this was the only young cluster
discovered.  Moreover, its colors are very similar to the
integrated colors of the arc, providing more evidence for
association.  Other cluster candidates that were fainter than
our survey limit are more widely distributed, and may be 
associated with either the arc or with the central gas disk.

In Figure~\ref{bertelli} we compare the cluster's {\it UBV} colors to the
single stellar population evolutionary tracks of Bertelli et al.\ (1994).  
Modulo any internal reddening, we obtain an age of
$\sim350$~Myr, possibly with a slightly sub-solar metallicity.  
The strong Balmer absorption lines evident in the star cluster's spectrum
(Figure~\ref{clusterfig}) are also age diagnostics.  We measured
the equivalent widths of the Balmer lines from H$\beta$ to H9 using the
line and continuum windows defined in Table~3 of Gonz{\'a}lez Delgado \&
Leitherer (1999).  The results are listed in Table~\ref{balmer}.  
When we compared these line widths to instantaneous
burst models at metallicities of solar and 5\% solar 
(Gonz{\'a}lez Delgado, Leitherer, \&
Heckman 1999), we find that they are consistent with those for a few
hundred Myr old stellar population.  Unfortunately, these
equivalent widths are sensitive to the continuum fit, and 
have relatively large errors of approximately $\pm2$\AA. 
Determining a more precise age and metallicity 
will require a spectrum with higher
signal-to-noise.  Nevertheless, the range of ages allowed by the
spectroscopy is consistent with the previously derived photometric age.

Could this cluster be a proto-globular cluster?  
With $M_V=-8.8$ and (\bv)$_0 = 0.24$, this cluster is
similar to the young massive clusters in spiral galaxies
studied by Larsen \& Richtler (1999).
When we compared its $M_V$ and age to the fading tracks of
Girardi \& Bica (1993), assuming a Salpeter IMF and a metallicity
of Z=0.004, we find that the cluster had an
initial mass of approximately $10^5 M_{\sun}$, and that in
15 Gyr it will fade to $M_V\sim -7$.  This makes it close in mass to the
mean mass of the Milky Way globular cluster system, which is
$\bar{M}=1.9\times10^5M_{\sun}$ (Mandushev, Staneva \& Spasova 1991).
The blue cluster's appearance is as compact as that of old \cena\ 
globular clusters, and while it is difficult to predict the disruptive
forces that it will experience, it is likely that this
cluster will become indistinguishable from the low mass 
globular clusters in the \cena\ halo as it ages.  

In addition to this spectroscopically confirmed cluster, there are
a number of other blue objects in the vicinity of the arc.  In
Figure~\ref{clusterpos} we also show the locations of all discrete
sources that have apparent magnitudes $18<V_0<23$ and colors bluer than 
(\bv)$ < 0.3$.  Many of these objects lie along the blue arc, while
others are probably associated with the central gas disk.
The blue color of the arc is similar to that of an A-type star.  A stellar
population that is younger than $\sim 50$ Myr would contain individual
A-supergiants with $-9<M_V<-5$, an absolute magnitude range that
corresponds to the apparent magnitude range of our ``blue objects''.  
While it is possible that some of the
blue objects are supergiant stars in \cena, preliminary imaging with the
Very Large Telescope/Kueyen taken in $0\farcs6$ seeing show that all of
the blue objects that lie on the arc are resolved.  Therefore, it is
likely that they are fainter star clusters rather than A-supergiants.
Further observations to confirm the nature of these
objects will be obtained and discussed in a future paper.

The combined light of the unresolved and discrete blue light in the
brightest NW portion of the arc is listed in Table~\ref{arcphot}.
The total luminosity of this arc is $L_B = 4.8\times10^6L_{\sun}$.

\section{The Tidal Disruption of an Accreted Dwarf Irregular Galaxy?}

Previous studies of fine structure in \cena\ show that the galaxy has a 
rich shell structure that likely is a result of the recent merger event
(MQG83).  Could the blue arc be related to the shells?  
MQG83, in their shell catalog, labeled the brightest portion of the arc
as ``Feature~8'' and described it as a ``filament''.
With the benefit of having multicolor CCD observations, we 
compared the unsharp masked image (Figure~\ref{shells})
to the color map (Figure~\ref{colormap}).  The shells, which are
ubiquitous within the inner few effective radii, differ
from the blue arc in their color.  Outside of the central dust
obscuration and the blue arc, the color structure of the galaxy is
relatively smooth, showing little of the fine
structure seen in Figure~\ref{shells}.  This color difference between
the arc and the shells is also visible in the
color image in Figure~\ref{ahe_full}.  Moreover, the arc is not concentric
along the major axis as are most of the known shells.  The one exception
is a sharp blue feature in the SW corner of the image that
corresponds to an off-center shell.

While it is difficult to rule out a shell-like origin for the arc, our
color maps allow us to distinguish
it clearly from the surrounding structure.  It is possible that this is a young
shell---one might expect inner shells to be bluer because the orbital
timescale is short.  However, there are many shells inside of the arc
that are not as blue, implying that the dynamical time is not likely to be
the only explanation for the color difference.  With its A-star colors and
young star clusters, the arc has a different 
stellar population from the other shells, and from a 
canonical elliptical galaxy.

We propose that the arc is the product of a recent tidal disruption.
The morphology, brightness, and associated stellar populations of the arc
suggest a stream of young stars following an orbit through the 
\cena\ potential.  It is unlike
previously detected halo streams---e.g.\ NGC~5907 and those associated
with the Sgr dwarf in the Milky Way---because it is composed primarily of
young stars.  Dwarf irregular (dI) galaxies, however, are gas-rich and
contain young stellar populations.  Given the large numbers of dIs in
the Centaurus Group (C\^{o}t\'{e} \etal 1997; Banks \etal 1999), 
it is possible that one has
been accreted by \cena, perhaps in association with the larger merger
event that created the shells and central gas disk.  An infalling
dwarf would be tidally disrupted, creating a stream of young stars in the
halo.  Many of these stars may have been formed recently, in an event 
triggered by the encounter.  

To create a young tidal stream in this fashion, the dI would need to
be massive and compact enough to retain gas as it passed close to the center of
\cena, but not so massive that it would survive the encounter largely
intact.  Would we expect this to be the case for an infalling dI?
The tidal limit, $r_T$, for a dwarf galaxy orbiting a massive host is
given by (Lang 1974; von Hoerner 1957)

\begin{equation}
r_T = R\ (\frac{m}{3M})^{1/3}
\end{equation}

\noindent where $m$ is the mass of the dwarf, $M$ is the mass of the
host galaxy, and R is the distance between galaxies.  Using the mass
profile derived by Hui \etal (1995), we adopt masses for 
\cena\ within 25, 8, and 2~kpc as $3.1\times10^{11}$,
$1.1\times10^{11}$, and $0.25\times10^{11}\ M_{\sun}$, respectively.
We assume a dwarf galaxy mass of $10^7\ M_{\sun}$.  
For a dwarf at these distances $R$, the tidal limit $r_T$ is
550, 250, and 100~pc.  The ``core'' radii of Local Group dwarf galaxies 
ranges from 95~pc (DDO 210) to 710~pc (WLM) (Mateo \etal 1998).  By this
measure, it is possible that a dI galaxy similar to those found in
the Local Group could retain much of its gas until it was well into the
main body of \cena.  Once the dI reaches the center, though, it is
likely to be severely disrupted by tidal forces.  The star formation
episode that accompanies this tidal interaction may also have a role in
disrupting the ISM of the dwarf.  How much of the galaxy
core remained concentrated would depend on its initial mass and density,
as well as the time since the encounter.  
While we do not detect such an overdensity along the orbit of the tidal
stream, the extinction in \cena's center could be hiding other parts of
the disrupted galaxy.

When compared to the dwarf galaxies in the Local Group,
the colors and luminosity of the stream are most similar to those of
Leo~A and SagDIG (Mateo 1998).  However, because our measured luminosity
for the arc does not necessarily include the entire content of the
progenitor, we must treat this measured value as a lower limit. 
The infalling galaxy may originally have been more extensive and luminous.

If the arc did originate in a minor gas-rich merger, we can use
its size to infer its mass and a timescale for the event.
JSB01 provide a useful semi-analytic formulation for the behavior of 
tidally disrupted stellar streams in a galaxy halo.  Because
it is impossible to know 
the true three-dimensional geometry of the arc, we made the simplifying
assumption that it is in the plane of the sky.  This allowed us to place
a lower limit on the disruption timescale and make an estimate of the
total mass.

JSB01 represent the time since disruption ($t$) as

\begin{equation}
t = 0.01\Psi (\frac{R}{w}) (\frac{R_{circ}}{10 {\rm kpc}}) 
(\frac{200 {\rm km/s}}{v_{circ}}) {\rm \ Gyr}
\end{equation}

\noindent where $w$ is the width of the streamer at radius $R$, $\Psi$ is the
angular length of the streamer, and $R_{circ}$ is the radius of a
circular orbit with the same energy as the true orbit.  The width of the
streamer is 490~pc at a projected galactocentric distance of 8~kpc.  It
is possible to trace
almost half of a complete ellipse in the ($B$--$R$) color map, so we
take a value of $\pi$ for $\Psi$.  The circular velocity at
8~kpc is 250~km/s (Hui \etal 1995), and we assume that $R_{circ}$
is approximately half way between the apocenter and pericenter of the
orbit, a value of 5.8~kpc.  This provides a lower limit on the time
since disruption, as there may be additional parts of the stream that
are not visible.  Using this formulation, we find that the time since
disruption is at least 240~Myr.

The one confirmed star cluster has an age that is consistent with this
disruption timescale.  By comparing the integrated color of the arc to
the Bertelli \etal (1994) evolutionary tracks in 
Figure~\ref{bertelli} we derive a luminosity-weighted, single
stellar population age of $\sim 200$ Myr.  The agreement of these three ages 
suggest a scenario where the stars in the arc
formed during a tidally triggered event as their parent body passed
through the potential of \cena.  

JSB01 also describe the mass $m$ of a stellar streamer as

\begin{equation}
m = 10^{11} (\frac{w}{R})^3 (\frac{R_p}{{\rm 10 kpc}}) (\frac{v_{circ}}{{\rm
200 km/s}})^2 M_{\sun}
\end{equation}

\noindent where $R_p$ is the pericentric distance of the orbit, and all
the other quantities are the same as above.  Modulo inclination effects
that change the values of $R$ and $R_p$, we derive a mass of
$7.7\times10^6 M_{\sun}$ for the blue tidal stream.
Using the luminosity measured in section~\ref{photometry}, this gives a
mass-to-light ratio of $M/L_B = 2.5$.
The total luminosity of the arc is likely to be underestimated because
of extinction, and because our measured value 
is only for the brightest quarter of the arc.
However, this is still a reasonable estimate because stars in a stream
will pile up at apocenter rather than evenly distribute themselves
through the orbit.  The resulting $M/L$ value is very similar to what is
measured in Local Group dIs (e.g. Mateo 1998).

Published \ion{H}{1} observations in the vicinity of the tidal stream allow
us to place more constraints on its nature.
A survey for \ion{H}{1} in \cena\ by Schiminovich et al.\ (1994) revealed
shell-like structures that they associate with the optical shells.
While their NW shell is close to the apocenter of the stream, 
there is no \ion{H}{1} detected along the tidal stream.
Likewise, Charmandaris, Combes, \& van der Hulst (2000) detected CO
in the \ion{H}{1} shells, but none in the vicinity of the tidal stream. 
In fact, their ``S4'' pointing was very close to the apocenter of the
stream, and was a non-detection.  The lack of neutral atomic or
molecular gas is not entirely unexpected, as we see neither a young 
($< 10$ Myr) population of massive stars nor any obvious \ion{H}{2}
regions that would be the signature of
ongoing star formation.  This is in contrast to the AGN 
jet-induced star formation
currently ongoing in the NE halo, where stars have ages of 1--15~Myr,
and the spatial groupings are much looser (Fassett \& Graham 2000).

Recent X-ray observations by Chandra also show diffuse structure in the
region of the tidal stream (Karovska \etal 2002).  It is also possible that
interactions with the hot halo gas may have affected the stream's formation or
evolution.

\section{Conclusions}

The detection of tidal debris streams in the halos of nearby galaxies is
a burgeoning field that promises to shed new light on galaxy building.
Using broadband optical color maps, we have identified one of the first
trails of {\it young} stellar debris---a young blue
tidal stream in the halo of the nearest giant elliptical galaxy, \cena.
Associated with this arc are numerous blue star clusters, one of which
is both spectroscopically confirmed and massive enough to be a young
globular cluster.  The mean age of the unresolved blue
light, age of the star cluster, and dynamical disruption 
timescale all have values of 200--400~Myr.
We propose that this stream of young stars was formed when
a dwarf irregular galaxy, or similar sized gas fragment, underwent a
tidally triggered star formation episode as it fell into 
\cena\ and was disrupted $\sim 300$~Myr ago.  
Non-detections to date of neutral or molecular gas in the stream 
is consistent with
the lack of obvious OB associations or resolved \ion{H}{2} regions,
implying that the star formation is no longer ongoing.   

The larger merger event that formed
the central gas disk has a wide range of published ages, from 200~Myr
(Quillen \etal 1993) to 750~Myr (Sparke 1996).  While it is strongly
possible that the formation of the tidal stream occurred in tandem with
the larger merger, perhaps as an infalling satellite of the larger galaxy, we
emphasize that the stream's morphology, stellar age, and gas content
now makes it distinct from both the central disk and the jet-induced
star formation in the NE halo.

The stars and star clusters from this tidal stream will eventually
disperse into the main 
body of \cena, suggesting that the late infall of gas-rich dwarf
galaxies may play an important a role in the building of stellar halos.  
Future spectroscopic observations of the blue star clusters
will provide valuable information on the stream's metallicity
distribution and kinematic structure.

\acknowledgments

EWP acknowledges support from NSF grant AST-0098566.  HCF acknowledges
support from NASA contract NAS5-32865 and NASA grant NAG5-7697.  We
thank the staff at CTIO for their invaluable help during our observing runs.
We also thank David Malin for making available to us his deep
photographic prints of \cena.
This research has made use of the NASA/IPAC Extragalactic Database (NED)
which is operated by the Jet Propulsion Laboratory, California Institute
of Technology, under contract with the National Aeronautics and Space
Administration.




\clearpage















\begin{deluxetable}{lrrrrrcccc}
\tabletypesize{\scriptsize}
\tablecaption{Photometry of Arc and Star Cluster \label{arcphot}}
\tablewidth{0pt}
\tablehead{
\colhead{Filter} & \colhead{$U_0$}   & \colhead{$B_0$}   &
\colhead{$V_0$} & \colhead{$R_0$} &
\colhead{$I_0$}  & \colhead{(\ub)$_0$} & \colhead{(\bv)$_0$} &
\colhead{(\vr)$_0$} & \colhead{(\vi)$_0$}
}
\startdata
Diffuse$^{\rm a}$ & 16.60 & 16.70 & 16.59 & 16.39 & & & & &  \\
Star Cluster & 19.12 & 19.15 & 18.91 & 18.68 & 18.42 & $-$0.04 &  0.24 &
0.23 & 0.49 \\
Total$^{\rm b}$ & 16.40 & 16.50 & 16.37 & 16.18 & & $-$0.10 & 0.13 &
0.19 & \\
Total(abs)$^{\rm c}$ & $-$11.32 & $-$11.22 & $-$11.35 & $-$11.54 & & & &
 & \\
\enddata

\tablenotetext{a}{Photometry for the diffuse component of the arc
(discrete sources masked)}
\tablenotetext{b}{Photometry for arc that includes blue discrete sources}
\tablenotetext{c}{Absolute magnitudes assuming $m-M = 27.72$}

\tablecomments{All photometry is corrected for Galactic foreground
reddening using the extinction maps of Schlegel, Finkbeiner, \& Davis
(1998).}

\end{deluxetable}



\begin{deluxetable}{ccccc}
\tabletypesize{\scriptsize}
\tablecaption{Equivalent Widths\tablenotemark{a} of Balmer Series 
for Blue Star Cluster \label{balmer}} 
\tablewidth{0pt}
\tablehead{
\colhead{H$\beta$ (\AA)} & \colhead{H$\gamma$ (\AA)} &
\colhead{H$\delta$ (\AA)} & \colhead{H8 (\AA)} &
\colhead{H9 (\AA)}
}
\startdata
11 & 14 & 13 & 13 & 10 \\
\enddata


\tablenotetext{a}{Measured using polynomial fit to continuum windows on
non-flux calibrated spectrum.  Errors are aproximately $\pm2$~\AA.}


\end{deluxetable}

\end{document}